\documentclass[final,5p,times,twocolumn]{elsarticle}
\usepackage[utf8]{inputenc}
\usepackage{amsmath,amssymb,amsfonts}
\usepackage{graphicx}
\usepackage{geometry}
\usepackage{enumitem}
\journal{Physics Letters B}

\begin{document}
	
	\begin{frontmatter}

\title{Fractal dimensions at the femtometer scale explain the deuteron without tensor force}
\author{Lorenzo Fortunato}
\affiliation{organization={Dipartimento di Fisica e Astronomia ``G. Galilei'', Universit\`a di Padova and INFN-Padova},
	city={Padova},
	country={Italy}}

	\begin{abstract}
	The deuteron serves as the ultimate touchstone for models of the nucleon-nucleon interaction. From simple experimental facts and quantum mechanical models, it has been deduced that the p-n potential must contain a tensor term, attributed to spin-spin interactions, that admixes the monopole and quadrupole components. Here we show that a consistent picture of the deuteron that explains energy, S-D wave mixing, magnetic dipole moment and electric quadrupole moments, can be achieved even without any explicit tensor potential by admitting a fractal distortion of the Euclidean space at the femtometer scale ($D =2+q \approx 3.014$). Using Tarasov's multi-fractional calculus,  the anisotropic part of the Laplace-Beltrami operator in fractal dimensions acts as an effective tensor coupling, providing all the salient features of the deuteron.
\end{abstract}

\end{frontmatter}

	\section{Introduction}
	In standard nuclear physics, the deuteron is modeled as a quantum system bound by a central nuclear force plus the nuclear tensor force introduced as a potential energy operator (\({\hat S}_{12}\)) that is modeled as a magnetic-dipole-like non-central interaction dependent on the alignment of the proton and neutron spins relative to the axis connecting them.
	This term is necessary to account for S-D wave mixing and the magnetic dipole and electric quadrupole moments.	
	
	Here a new perspective is adopted, that replaces the non-central tensor potential entirely by changing the underlying space geometry: the deuteron is modeled as a bound state embedded in a continuously deformed fractal space of dimensions $\approx 3.014123$. The central force is kept as a simple and spherical (square well for the sake of simplicity), but the mixing of angular momenta is attributed directly to the effect of a non-integer geometric manifold of the warped space at the femtometer scale.
	This derivation elegantly demonstrates that \(D=3\) is a unique symmetry point where angular momentum sectors are perfectly decoupled. Stepping away to \(D \gtrsim 3\) acts as a mathematical "torque" or "warp" that forces the S-state to mix with the D-state.
	
	While the empty space has been constrained by cosmologists to have a number of dimensions equal to 3 with high precision \cite{Car}, within the dense, highly packed interior of a nucleus, the concept of an "effective local fractal dimension" generated by the complex web of gluon and meson exchanges is an area of active interest.
	The notion that strong nuclear interactions can manifest as an effective distortion of local Euclidean space geometry at the femtometer scale has some distinguished precedent. Early ``strong gravity'' formulations by Salam and Sivaram~\cite{SalSiv} and geometric hadron models by Recami and Castorina~\cite{RecCas} proposed that massive spin-$2$ meson exchanges induce an effective short-range metric curvature inside hadronic matter. More recently field theories in fractional spaces developed by Calcagni~\cite{Cal} have shown that gauge-boson fluctuations and non-differentiable quantum paths naturally cause the effective Hausdorff and spectral dimensions of space to depart  from $D=3$ at short distances. Building on Tarasov's multi-fractional vector calculus~\cite{Tar}, we demonstrate here that an anisotropic axial deformation $D = 2 + q \gtrsim 3$ is sufficient to reproduce the crucial static properties of the deuteron---namely $S$-$D$ state mixing, the electric quadrupole moment, and the magnetic dipole moment---without introducing an explicit tensor potential. Of course, as it is well-known from meson theory and chiral EFT, pion exchange naturally generates a tensor term $\hat S_{12}$, therefore we do not intend to cancel it completely, but just to show that this bold alternative makes perfect sense and can compete with tensor interactions. 
	
	\section{General Tarasov Framework ($\alpha_1, \alpha_2, \alpha_3$)}
	In Tarasov's multi-fractional space \cite{Tar}, global spherical symmetry is replaced by directional vector calculus. A 3-dimensional coordinate system is generalized such that each independent Cartesian axis behaves as its own non-integer entity carrying a continuous dimension parameter $\alpha_k > 0$. The total dimension of the space is given by the sum:
	\begin{equation}
		D = \alpha_1 + \alpha_2 + \alpha_3
	\end{equation}
	The total volume element $dV_D$ is constructed as a product of independent, directional fractional measures:
	\begin{equation}
		dV_D(x,y,z) = d\mu_{\alpha_1}(x) \, d\mu_{\alpha_2}(y) \, d\mu_{\alpha_3}(z)
	\end{equation}
	where each directional measure scales as:
	\begin{equation}
		d\mu_{\alpha_k}(x_k) = \frac{\pi^{\alpha_k/2}}{\Gamma(\alpha_k/2)} |x_k|^{\alpha_k-1} dx_k
	\end{equation}
	
	\subsection{Anisotropic Derivatives, Gradient, and Cartesian Laplacian}
	The directional derivatives are defined as the operational inverse to the integration weights.
	
	For the $k$-th coordinate axis, the Tarasov directional derivative operator $\mathbb{D}_k^{(\alpha_k)}$ acting on a wave function $\psi$ is defined as:
	\begin{equation}
		\mathbb{D}_k^{(\alpha_k)} \psi = \frac{\Gamma(\alpha_k/2)}{\pi^{\alpha_k/2}} |x_k|^{1 - \alpha_k} \frac{\partial \psi}{\partial x_k}
	\end{equation}
	
	Grouping these directional components into a unified vector framework defines the anisotropic gradient $\mathbf{\nabla}_D$:
	\begin{equation}
		\mathbf{\nabla}_D = \sum_{k=1}^3 \mathbf{e}_k \, \mathbb{D}_k^{(\alpha_k)}
	\end{equation}
	
	The kinetic energy operator is proportional to the anisotropic Laplacian $\Delta_D$, defined via the vector dot product $\mathbf{\nabla}_D \cdot \mathbf{\nabla}_D$:
\begin{equation}
	\Delta_D = \sum_{k=1}^3 \left( \frac{\Gamma(\alpha_k/2)}{\pi^{\alpha_k/2}} \right)^2 |x_k|^{1 - \alpha_k} \frac{\partial}{\partial x_k} \left( |x_k|^{1 - \alpha_k} \frac{\partial}{\partial x_k} \right)
\end{equation}
	Expanding the derivatives explicitly in Cartesian coordinates yields:
\begin{equation}
	\Delta_D = \sum_{k=1}^3 A_k \Biggl[ |x_k|^{2 - 2\alpha_k} \frac{\partial^2}{\partial x_k^2} + (1 - \alpha_k)|x_k|^{1 - 2\alpha_k} \operatorname{sgn}(x_k) \frac{\partial}{\partial x_k} \Biggr]
\end{equation}
	where $A_k = \left( \frac{\Gamma(\alpha_k/2)}{\pi^{\alpha_k/2}} \right)^2$ gathers the directional normalization constants.
	
	\subsection{Simplification to axial deformation ($\alpha_3 = q$)}
	We restrict the non-integer behavior strictly to the polar $z$-axis, taken as the spin quantization axis, leaving the transverse directions $x$ and $y$ dimensions as in standard 2D Euclidean space:
	\begin{equation}
		\alpha_1 = 1, \quad \alpha_2 = 1, \quad \alpha_3 = q
	\end{equation}
	Since $\frac{\pi^{1/2}}{\Gamma(1/2)} = 1$, the normalization constants simplify to $A_1 = A_2 = 1$ and $A_3 = A_q = \left(\frac{\Gamma(q/2)}{\pi^{q/2}}\right)^2$.
	Physically, this axial anisotropy should not be viewed in the empty laboratory space, which would violate $\mathrm{SO}(3)$ rotational invariance, but rather as an \emph{intrinsic} geometric deformation aligned dynamically with the deuteron's total spin quantization axis $\mathbf{J}$ (much like intrinsic quadrupole deformations in the Bohr--Mottelson collective model). Projecting onto the lab-frame preserves the degeneracy of the magnetic substates ($M_J = 0, \pm 1$) in the absence of external fields.

	The modified volume element contracts to a sort of cylindrical geometry along the $z$-axis:
	\begin{equation}
		dV_q = \left( \frac{\pi^{q/2}}{\Gamma(q/2)} |z|^{q-1} dz \right) dx \, dy
	\end{equation}
	Converting to spherical coordinates via $z = r\cos\theta$, $x = r\sin\theta\cos\phi$, and $y = r\sin\theta\sin\phi$, the volume element becomes:
	\begin{equation}
		dV_q = A_q^{-1/2} \, r^{q+1} \sin\theta \, |\cos\theta|^{q-1} \, dr \, d\theta \, d\phi \label{v_elem}
	\end{equation}
	
	The axially deformed Laplacian simplifies to:
	\begin{equation}
		\begin{aligned}
			&\Delta_D= \\
			&= \frac{\partial^2}{\partial x^2} + \frac{\partial^2}{\partial y^2} + A_q \Bigg[ |z|^{2-2q} \frac{\partial^2}{\partial z^2} + (1 - q)|z|^{1-2q} \operatorname{sgn}(z) \frac{\partial}{\partial z} \Bigg] \\
			&= \Delta_3 + \underbrace{\left( A_q |z|^{2-2q} - 1 \right) \frac{\partial^2}{\partial z^2}A_q +(1 - q)|z|^{1-2q} \operatorname{sgn}(z) \frac{\partial}{\partial z}}_{\Delta_{\text{aniso}}} 
		\end{aligned}
	\end{equation}
	where the isotropic-3D and anisotropic (still hermitian) part, $\Delta_3$ has been separated out.

\section{Coupled-Channel Generalized Equations}
\label{sec:coupled}

The problem we face is analogous to the Rarita--Schwinger formulation \cite{RaSc}, with the difference that the phenomenological tensor potential $V_T$ is entirely absent. Instead, new kinetic operators arise from the anisotropic Laplacian $\Delta_{\text{aniso}}$, which not only couple the $S$- and $D$-wave channels off-diagonal, but also contribute diagonal kinetic energy shift terms in both the $S$-$S$ and $D$-$D$ sectors. Expanding the ground-state deuteron wavefunction into its $L=0$ and $L=2$ spin-angular components gives:
\begin{equation}
		\Psi_{1, M_J}(\mathbf{r}) = \frac{u(r)}{r} \mathcal{Y}_{0,1,1}^{M_J}(\theta, \phi) + \frac{w(r)}{r} \mathcal{Y}_{2,1,1}^{M_J}(\theta, \phi) \,,
\end{equation}
where the spin-angle harmonics are constructed using standard Clebsch--Gordan coupling,
\begin{equation}
	\begin{aligned}
		\mathcal{Y}_{LSJ}^{M_J}(\theta, \phi) &= \sum_{M_L, M_S} \langle L, M_L; S, M_S \mid J, M_J \rangle \\
		&\qquad\quad \times Y_{L, M_L}(\theta, \phi) \, \chi_{S, M_S} \,,
	\end{aligned}
\end{equation}
and normalized with respect to the Tarasov angular measure $d\Omega_q = A_q^{-1/2} \sin\theta \, \lvert\cos\theta\rvert^{q-1} \, d\theta \, d\phi$.

Because the polar weight $\lvert\cos\theta\rvert^{q-1}$ breaks strict Euclidean orthogonality between $L=0$ and $L=2$ for $q \neq 1$, projecting the Schr\"odinger equation $(\hat{H}_D - E)\Psi_{1, M_J} = 0$ onto the angular basis yields a $2 \times 2$ generalized coupled-channel system of differential equations:
\begin{equation}
	\begin{aligned}
		&\begin{pmatrix} 
			\hat{H}_{00} & \hat{T}_{02} + \mathcal{O}_{02} V_C \\[6pt] 
			\hat{T}_{20} + \mathcal{O}_{20} V_C & \hat{H}_{22}
		\end{pmatrix}
		\begin{pmatrix}
			u(r) \\
			w(r)
		\end{pmatrix} \\
		&\qquad = E
		\begin{pmatrix}
			1 & \mathcal{O}_{02} \\[4pt]
			\mathcal{O}_{20} & 1
		\end{pmatrix}
		\begin{pmatrix}
			u(r) \\
			w(r)
		\end{pmatrix} \,,
	\end{aligned}
\end{equation}
where $\hat{H}_{00}(r) = -\frac{\hbar^2}{2\mu}\frac{d^2}{dr^2} + V_C(r) + \hat{T}_{00}(r)$, $\hat{H}_{22}(r) = -\frac{\hbar^2}{2\mu}\left(\frac{d^2}{dr^2} - \frac{6}{r^2}\right) + V_C(r) + \hat{T}_{22}(r)$, $\hat{T}_{LL'}(r)$ denote the radial differential operators induced by $-\frac{\hbar^2}{2\mu}\Delta_{\text{aniso}}$, and $\mathcal{O}_{02}(q) = \mathcal{O}_{20}^*(q) = \int \mathcal{Y}_{0,1,1}^{M_J\,\dagger} \mathcal{Y}_{2,1,1}^{M_J} \, d\Omega_q$ is the angular overlap due to fractal measure. Written explicitly, the coupled radial equations read:
\begin{align}
	\left[ -\frac{\hbar^2}{2\mu} \frac{d^2}{dr^2} + V_C(r) + \hat{T}_{00}(r) - E \right] u(r) &\nonumber\\
	+ \left[ \hat{T}_{02}(r) + \mathcal{O}_{02}(q)\big(V_C(r) - E\big) \right] w(r) &= 0 \,, \\[6pt]
	\left[ -\frac{\hbar^2}{2\mu} \left( \frac{d^2}{dr^2} - \frac{6}{r^2} \right) + V_C(r) + \hat{T}_{22}(r) - E \right] w(r) &\nonumber\\
	+ \left[ \hat{T}_{20}(r) + \mathcal{O}_{20}(q)\big(V_C(r) - E\big) \right] u(r) &= 0 \,.
\end{align}
At the Euclidean symmetry point $q = 1$, both $\mathcal{O}_{02}(1) = 0$ and $\hat{T}_{LL'}(r) = 0$, restoring the standard uncoupled radial equations. 

\section{Matrix Element $\langle \psi_S | \hat{T} | \psi_D \rangle$ }
We evaluate the mixing matrix element of the kinetic energy operator $\hat{T} = -\frac{\hbar^2}{2\mu}\Delta_D$ between the lowest spherical $S$-state and the polar quadrupole $D$-state ($Y_2^0$):
\begin{equation}
	\begin{aligned}
		\psi_S(\mathbf{r}) &= \frac{1}{\sqrt{4\pi}} R_S(r), \\
		\psi_D(\mathbf{r}) &= \sqrt{\frac{5}{16\pi}} R_D(r) (3\cos^2\theta - 1)
	\end{aligned}
\end{equation}

where $R \approx 2.1 \text{ fm}$ represents the nuclear core radius and  $V_0 \simeq 33.72$ MeV. The S-wave is a standard piecewise function with $\kappa = \sqrt{2\mu E_B}/\hbar$, and $k_0 = \sqrt{2\mu(V_0 - E_B)}/\hbar$. For the D-wave(L=2) basis state, we adopt a normalized Sturmian trial function with proper exponential tail behaviour, $\propto e^{-\beta_Dr}$, where $\beta_D$ is the Sturmian range parameter.
Returning to the matrix element: 
\begin{equation}
	\begin{aligned}
		\langle \psi_D | \hat{T} | \psi_S \rangle &= -\frac{\hbar^2}{2m} \langle \psi_D | \Delta_D | \psi_S \rangle \\
		&= -\frac{\hbar^2}{2m} \langle \psi_D | \Delta_{\text{aniso}} | \psi_S \rangle 
	\end{aligned}
\end{equation}
because the $\Delta_3$ part does not mix different $L$'s.

\begin{table*}[!ht]
	\centering
	\small
	\setlength{\tabcolsep}{4pt}
	\caption{Self-consistent model parameters and predicted deuteron ground-state observables at $\beta_D = 0.33\text{ fm}^{-1}$ and $R_0 = 2.10\text{ fm}$, compared with experimental values and standard realistic nucleon-nucleon potential benchmarks.}
	\label{tab:deuteron_results}
	\renewcommand{\arraystretch}{1.2}
	\begin{tabular}{lccc}
		\hline\hline
		\textbf{Quantity} & \textbf{Unmixed ($q = 1$)} & \textbf{Deformed ($q = q^*$)} & \textbf{Expt. / Benchmark} \\
		\hline
		Central well depth $V_0^*$ (MeV) & $33.436$ & $33.436$ & $\sim 33.7$ (pure $S$-wave) \\
		Axial dimension parameter $q^*$ & $1.000000$ & $1.014123$ & --- \\
		Effective local dimension $D^* = 2 + q^*$ & $3.000000$ & $3.014123$ & $3.000$ (vacuum) \\
		Ground-state energy $E_{\text{gs}}$ (MeV) & $-2.1148$ & $-2.2250$ (fit) & $-2.2246$ \\
		Electric quadrupole moment $Q_d$ ($\text{fm}^2$) & $0.0000$ & $0.2859$ (fit) & $0.2859(3)$ \\
		$D$-state probability $P_D$ ($\%$) & $0.0000$ & $5.066$ & $4.5\text{–}5.8$ (NN models) \\
		Magnetic dipole moment $\mu_d$ ($\mu_N$) & $0.8798$ & $0.8509$ & $0.8574$ \\
		\hline\hline
	\end{tabular}
\end{table*}

\section{Generalized Two-State Mixing in Non-Orthogonal Basis}
\label{sec:mixing}

\begin{figure}[t]
	\centering
	\includegraphics[width=\columnwidth]{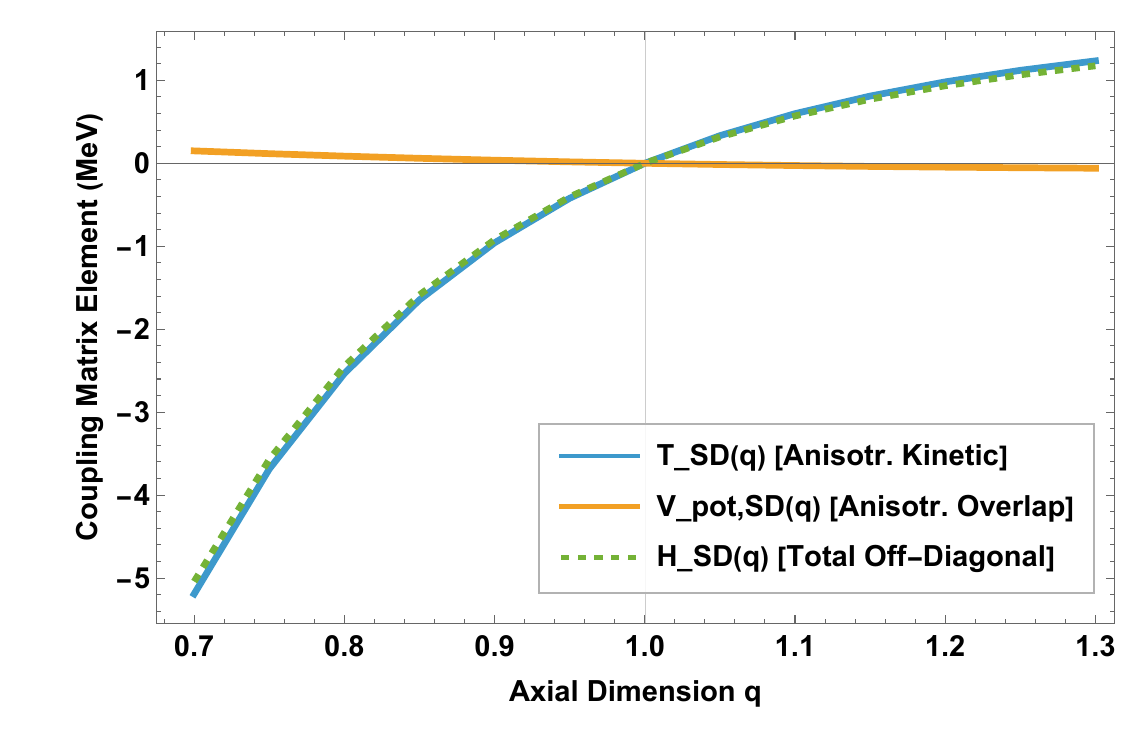}
	\includegraphics[width=\columnwidth]{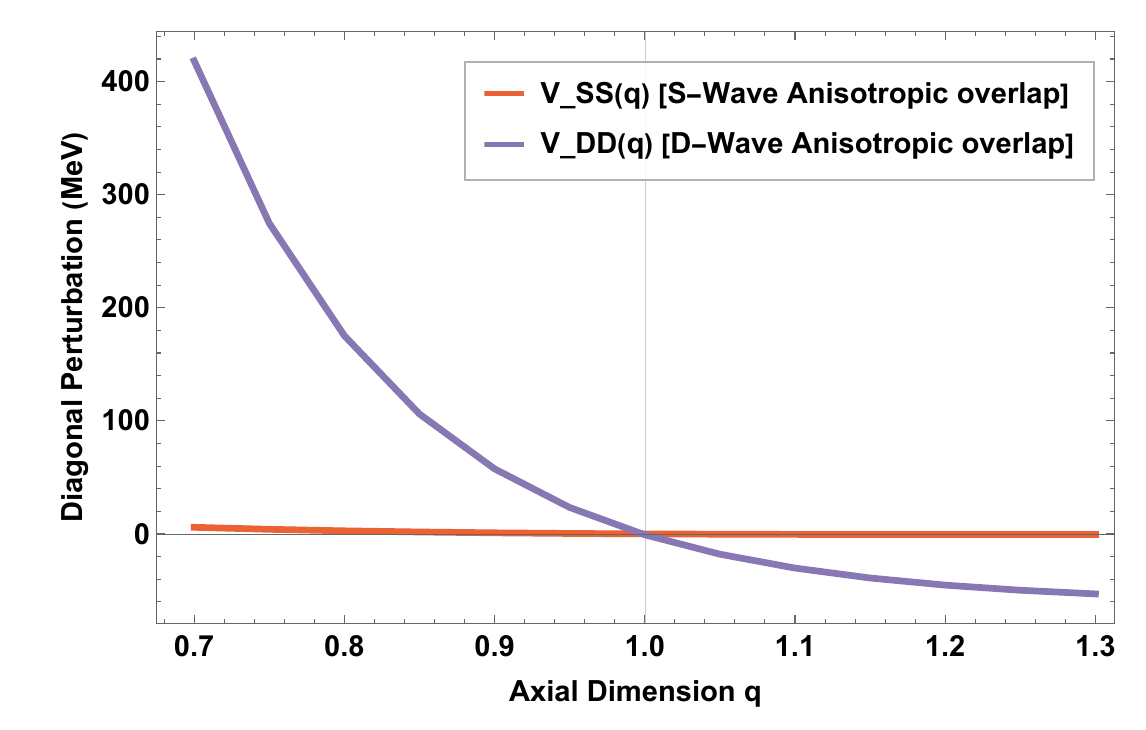}
	\caption{Matrix elements (coupling and overlap terms) as a function of $q$. Notice they cross zero at $q=1$.}
	\label{fig:couplings}
\end{figure}

While the basis states $\psi_S$ and $\psi_D$ are individually normalized in the fractional volume element ($\langle \psi_S \vert \psi_S \rangle_D = \langle \psi_D \vert \psi_D \rangle_D = 1$), the anisotropic polar weight $\lvert\cos\theta\rvert^{q-1}$ breaks spherical orthogonality whenever $q \neq 1$. Consequently, the basis states acquire a non-vanishing overlap integral,
\begin{equation}
	S_{SD}(q) = \langle \psi_D \vert \psi_S \rangle_D \neq 0 \quad (q \neq 1) \,,
\end{equation}
as well as a non-zero off-diagonal matrix element of the central square-well potential $V(r)$,
\begin{equation}
	\begin{aligned}
		V_{\mathrm{pot},SD}(q) &= \langle \psi_D \vert V(r) \vert \psi_S \rangle_D \\
		&= -V_0 \int_{r \le R} \psi_D^*(\mathbf{r}) \psi_S(\mathbf{r}) \, dV_D \,.
	\end{aligned}
\end{equation}
Both $S_{SD}(q)$ and $V_{\mathrm{pot},SD}(q)$ vanish identically in the isotropic limit $q = 1$, as seen in Fig. \ref{fig:couplings}.
Expanding the coupled deuteron ground state as $\Psi_{1,1} = c_S \psi_S + c_D \psi_D$, the normalization condition becomes
\begin{equation}
	\langle \Psi_{1,1} \vert \Psi_{1,1} \rangle_D = c_S^2 + c_D^2 + 2 c_S c_D S_{SD}(q) = 1 \,.
\end{equation}
Minimizing the energy expectation value leads to the $2 \times 2$ generalized eigenvalue problem $(H - E S)\mathbf{c} = 0$:
\begin{equation}
	\begin{pmatrix}
		H_{SS} - E & H_{SD} - E S_{SD} \\
		H_{SD} - E S_{SD} & H_{DD} - E
	\end{pmatrix}
	\begin{pmatrix} c_S \\ c_D \end{pmatrix} = 0 \,,
\end{equation}
where $H_{SS} = E_{S0} + V_{SS}$, $H_{DD} = E_{D0} + V_{DD}$, and $H_{SD}(q) = \langle \psi_D \vert \hat{T} \vert \psi_S \rangle_D + V_{\mathrm{pot},SD}(q)$. Fixing the lower eigenvalue to the experimental deuteron binding energy $E_- = E_B = -2.225\text{ MeV}$ yields the mixing ratio $r(q) = c_D / c_S$:
\begin{equation}
	r(q) = -\frac{H_{SD}(q) - E_B S_{SD}(q)}{H_{DD}(q) - E_B} \,,
\end{equation}
from which the normalized $D$-state probability is
\begin{equation}
	P_D(q) = c_D^2(q) = \frac{r^2(q)}{1 + 2 r(q) S_{SD}(q) + r^2(q)} \,.
\end{equation}

\section{Results and Discussion}
\label{sec:results}
\begin{figure*}[!h]
	\centering
	\includegraphics[width=\textwidth, clip=]{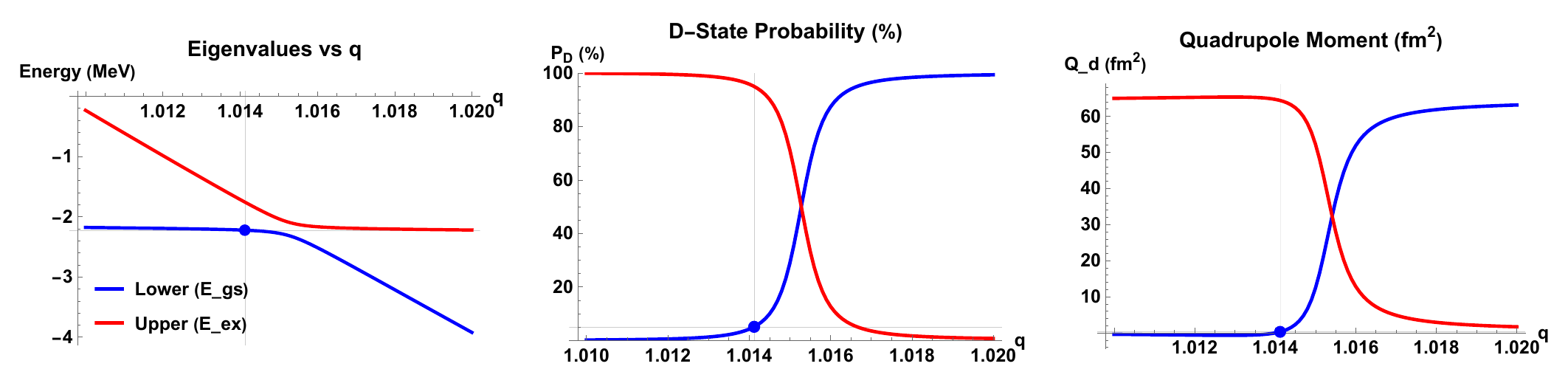}
	\caption{Determination of the value of q that simultaneously gives energy, probability and quadrupole moment.}
	\label{fig:results}
\end{figure*}
To test whether an axially deformed local space geometry can account for the static properties of the deuteron without an explicit tensor potential $\hat{S}_{12}$, we solve the $2 \times 2$ generalized eigenvalue problem $H(V_0, q)\mathbf{c} = E(V_0, q)S(q)\mathbf{c}$ numerically. We adopt standard physical constants ($\hbar c = 197.327\text{ MeV}\cdot\text{fm}$, $m_p c^2 = 938.272\text{ MeV}$, $m_n c^2 = 939.565\text{ MeV}$, giving a reduced mass $\mu c^2 = 469.459\text{ MeV}$ and $\hbar^2/(2\mu) = 41.471\text{ MeV}\cdot\text{fm}^2$) and fix the central spherical square-well radius to $R_0 = 2.10\text{ fm}$.

Typically, the central potential alone does not supply the full binding of the deuteron; rather, the $S\text{–}D$ coupling lowers the ground-state eigenvalue to the experimental value $E_B = -2.225\text{ MeV}$. Adopting the same self-consistent strategy here, we treat the central square-well depth $V_0$ and the axial fractional dimension $q$ as free geometric-potential parameters for a given $D$-wave Sturmian range $\beta_D$.

At the Euclidean symmetry point $q = 1$ ($D = 3$), all anisotropic kinetic matrix elements ($V_{SS}$, $V_{DD}$, $T_{SD}$) and the overlap $S_{SD}$ vanish identically, as seen in Fig. \ref{fig:couplings}, decoupling the $L=0$ and $L=2$ sectors so that $P_D(1) = 0$ and $Q_d(1) = 0$. Stepping $q$ away from unity induces an off-diagonal kinetic coupling $T_{SD}(q)$ dominated by the fractional Laplacian operator $\Delta_{\text{aniso}}$, while the central potential overlap $V_{\mathrm{pot},SD}(q) = -V_0 S_{SD}(q)$ remains small.

Setting the Sturmian range parameter to $\beta_D = 0.33\text{ fm}^{-1}$ and requiring the coupled ground-state eigenvalue and electric quadrupole moment to match experiment simultaneously,
\begin{equation}
	\begin{aligned}
		E_{\text{gs}}(V_0^*, q^*) &= -2.225\text{ MeV} \,, \\
		Q_d(V_0^*, q^*) &= 0.2859\text{ fm}^2 \,,
	\end{aligned}
\end{equation}
yields the self-consistent parameter set summarized in Table~\ref{tab:deuteron_results}.

\begin{figure*}[htbp]
	\centering
	\includegraphics[width=\textwidth]{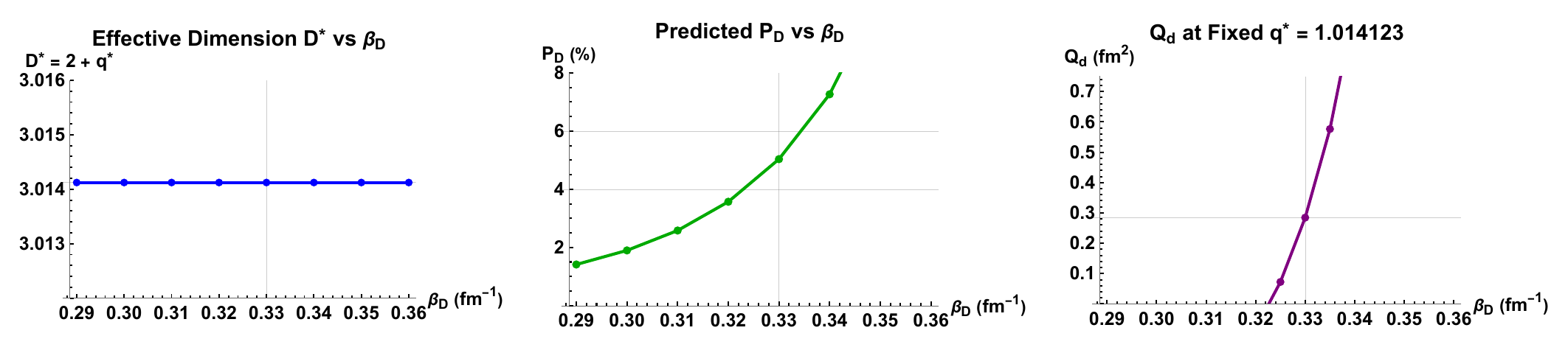}
	\caption{Sensitivity analysis across the Sturmian range parameter $\beta_D \in [0.29, 0.36]\text{ fm}^{-1}$: effective local dimension $D^* = 2 + q^*$ (left) and predicted $D$-state probability $P_D$ (center) when fitting $Q_d = 0.2859\text{ fm}^2$, alongside $Q_d(\beta_D)$ at fixed $q^* = 1.014123$ (right).}
	\label{fig:sensitivity}
\end{figure*}

A few interesting  physical features emerge from this solution:
\begin{itemize}[wide=0pt, leftmargin=0pt, itemsep=2pt, topsep=3pt]
	\item \textbf{Minimal dimensional departure:} Reproducing the experimental quadrupole moment requires an axial dimension parameter $q^* = 1.014123$, corresponding to an effective local fractal dimension $D^* = 3.014123$—a departure of less than $0.5\%$ from standard 3D Euclidean space. This mild distortion might be generated by the dense meson and gluon exchanges at the femtometer scale.
	\item \textbf{Simultaneous prediction of $P_D$ and $\mu_d$:} Once $(V_0^*, q^*)$ is fixed by $E_{\text{gs}}$ and $Q_d$, the $D$-state admixture is predicted without additional parameters to be $P_D = \lvert \langle \psi_D \vert \Psi \rangle_q \rvert^2 = 5.066\%$, well within the $4.5\%\text{–}5.8\%$ range characteristic of realistic meson-exchange potentials \cite{Mac}. Evaluating the impulse-approximation magnetic dipole moment,
	\begin{equation}
		\mu_d = (\mu_p + \mu_n) - \frac{3}{2}\left(\mu_p + \mu_n - \frac{1}{2}\right) P_D \,,
	\end{equation}
	with $\mu_p = 2.79285\,\mu_N$ and $\mu_n = -1.91304\,\mu_N$ yields $\mu_d = 0.8509\,\mu_N$, agreeing with the experimental value $0.8574\,\mu_N$ to within $0.75\%$ (with the small remaining discrepancy attributable to various corrections and to the simplicity of the model wavefunctions).
	\item \textbf{Self-similarity/ statistical Hausdorff dimensions:} The non-integer dimensions $q$ or $D$ need not necessarily be regular or self-similar (as in the Koch curve or Sierpinski gasket), they might also arise from statistical phenomena. In either interpretation, the notions of metric-space coverings and Hausdorff dimension remain mathematically well-defined.
\end{itemize}
At the intermediate nuclear binding scale (\(r \sim 2\text{ fm}\)) where meson-exchange clouds overlap between two nucleons, this fit finds a slight dimensional excess (\(D^* \approx 3.0141 > 3\)), reflecting a sort of effective proliferation of virtual meson/gluon degrees of freedom (a ``crumpled" or denser effective spatial volume) that generates the prolate quadrupole deformation (\(Q_d > 0\)) of the deuteron.

\textbf{Robustness.} To verify that our results do not rely on fine-tuning the Sturmian range parameter $\beta_D$, we repeat the self-consistent $(V_0^*, q^*)$ fit across the interval $\beta_D \in [0.29, 0.36]\text{ fm}^{-1}$ (corresponding to characteristic radial scales $1/\beta_D \approx 2.78\text{--}3.45\text{ fm}$). As shown in Fig.~\ref{fig:sensitivity} (left and center panels), the fitted effective local dimension remains remarkably stable across the entire range in $\beta_D$, staying confined within $3.0137 \le D^* \le 3.0149$ ($< 0.04\%$ relative variation), while $V_0^*$ remains nearly constant at $33.425\text{--}33.444\text{ MeV}$. Across the same window, the predicted $D$-state probability $P_D \in [4.05\%, 6.19\%]$ and magnetic dipole moment $\mu_d \in [0.845, 0.857]\,\mu_N$ remain consistent with realistic nucleon-nucleon models.

Conversely, holding the axial dimension fixed at $q^* = 1.014123$ while adjusting $V_0$ to preserve $E_{\text{gs}} = -2.225\text{ MeV}$ shows that $Q_d(\beta_D)$ varies quickly, hitting $0.2859\text{ fm}^2$ at a precise value (Fig.~\ref{fig:sensitivity}, right panel). Thus, while $Q_d$ sharply constrains the $D$-wave spatial scale to $\beta_D \approx 0.33\text{ fm}^{-1}$, the inferred space dimension $D^* \gtrsim 3$ is a quite insensitive feature of the model.

One last point worth mentioning is that we have adopted a basic model for the central potential (square well) and a more refined calculation (Woods-Saxon for example) might come up with a slightly different value of $q$, but we donot expect changes to the overall picture. Even reinstating a certain amount of tensor potential (but not as much as previously tought) that joins the effects of the space anisotropy might be a viable alternative.

\section{Conclusions}

In summary, it has been shown that, admitting the deuteron to live in a space with fractal dimensions (constrained to axial for simplicity) and using Tarasov's prescriptions for taking derivatives and gradient, leads to the non-trivial prediction that the static feature of this nucleus, i.e. the S-D wave mixing and the M1 and E2 moments, can be consistently obtained. The mixing is here induced by the anisotropic deformation of the Euclidean space, that is reflected in the kinetic term, rather than by the presence of a tensor potential. The main finding is that at $D = 2 + q \approx 3.014123$ all these data are simultaneously explained. Clearly it might be difficult to accept a non-integer dimension, but studies on fractals in condensed matter have shown without doubt that electronic wavefunctions adapt to empty or hollow systems (gaskets with $2<D<3$) \cite{Kemp, Eek}. Here, we are proposing that strong interactions might embroider (or dress) 3D Euclidean space at the femtometer scale, introducing an effective non-integer dimensionality slightly larger than 3. It remains to be seen how this might affect scattering properties.

\end{document}